\documentclass[aps,prd,eqsecnum,preprint,tightenlines,nofootinbib]{revtex4-1}
\usepackage{graphicx,latexsym}

\newcommand{\bea}{\begin{eqnarray}}
\newcommand{\eea}{\end{eqnarray}}
\newcommand{\be}{\begin{equation}}
\newcommand{\ee}{\end{equation}}
\newcommand{\ub}[1]{\underline{#1}}
\newcommand{\ob}[1]{\overline{#1}}
\newcommand{\Pminus}{{\cal P}^-}

\begin{document}

\title{Pauli--Villars regularization of non-Abelian gauge theories%
\footnote{Based on an invited talk presented at the
Lightcone 2015 workshop, Frascati, Italy, 
September 21-25, 2015.}
}

\author{J.R. Hiller}
\affiliation{Department of Physics and Astronomy\\
University of Minnesota-Duluth \\
Duluth, Minnesota 55812}

\date{\today}

\begin{abstract}
As an extension of earlier work on QED, we construct a 
BRST-invariant Lagrangian for SU(N) Yang-Mills theory
with fundamental matter,
regulated by the inclusion of massive Pauli-Villars (PV)
gluons and PV quarks.  The underlying gauge symmetry for massless PV
gluons is generalized to accommodate the PV-index-changing 
currents that are required by the regularization.  Auxiliary 
adjoint scalars are used, in a mechanism due to Stueckelberg, 
to attribute mass to the PV gluons and the PV quarks.  The addition of
Faddeev--Popov ghosts then establishes a residual BRST 
symmetry.  Although there are drawbacks to the approach, in particular
the computational load of a large number of PV fields
and a nonlocal interaction of the ghost fields, this
formulation could provide a foundation for renormalizable
nonperturbative solutions of light-front QCD in an arbitrary 
covariant gauge.
\end{abstract}

\maketitle

%%%%%%%%%%%%%%%%%%%%%%%%%%%%%%%%%%%%%%%%%%%%%%%%%%%%%
\section{Introduction} \label{sec:intro}
%%%%%%%%%%%%%%%%%%%%%%%%%%%%%%%%%%%%%%%%%%%%%%%%%%%%%

Nonperturbative Hamiltonian approaches to quantum field
theories~\cite{LFreview,Vary} in more than two dimensions require regularization.
The standard approach to regularization of non-Abelian
gauge theories is dimensional regularization~\cite{dimreg}, but this is 
inherently perturbative.  For Abelian theories,
Pauli--Villars (PV) regularization has proven useful~\cite{TwoPhoton,ArbGauge}.
However, the ordinary PV regularization of non-Abelian theories fails.
Gauge invariance is violated, blocking any hope of BRST invariance,
which confounds proofs of renormalizability~\cite{BRST}.
To implement a PV regularization~\cite{BRSTPVQCD}, we repair gauge invariance
and then break it properly with a gauge-fixing term.  Ghost
fields~\cite{FaddeevPopov} are added to build a BRST invariance.

The causes of gauge non-invariance in ordinary PV
regularization are the PV gluon mass, the non-degenerate
PV quark masses, and the PV-index changing currents for 
null interactions.  A mass is acceptable for PV photons
in an Abelian theory~\cite{MassiveVector}, but the approach
used does not generalize to non-Abelian theories.  Null
interactions (meaning interactions that involve only
null combinations of positive and negative metric fields)
are necessary to provide the subtractions that regulate
theory.  So, how can one restore gauge invariance?

The Higgs mechanism is, of course, one possible approach,
but investigations of this invariably found only
null combinations of gluon fields to be massless, rather
than a massless physical gluon.  There is, however, another
mechanism for generating masses~\cite{Stueckelberg,Stueckelberg2,Marnelius:1997rx}, by 
coupling the PV gluons and PV quarks to a PV scalar, 
with all removed from the spectrum in
the infinite-mass limit.  The mixing of currents can be
overcome by extension of the gauge transformation to
include mixing PV fields; the original gauge transformation
is obtained when the fields are removed from the spectrum.
The construction is done is such a way as to facilitate
quantization in an arbitrary covariant gauge, as has
been done previously for QED~\cite{ArbGauge}.  The
remaining gauge invariance can be tested by studying
the dependence of physical quantities on the gauge-fixing
parameter.

In the remainder of this paper, we present the PV-regulated
Lagrangian term by term, along with the extended gauge
transformation and the BRST transformations.  We then
discuss light-front quantization in an arbitrary covariant gauge
and the light-front coupled-cluster method~\cite{LFCClett} for solution of
the Hamiltonian eigenvalue problem in Fock space.

%%%%%%%%%%%%%%%%%%%%%%%%%%%%%%%%%%%%%%%%
\section{A Pauli--Villars Regulated Lagrangian}
%%%%%%%%%%%%%%%%%%%%%%%%%%%%%%%%%%%%%%%%

For the purposes of discussion, we divide the PV-regulated Lagrangian
into four terms
\be
{\cal L}={\cal L}_{\rm massless}+{\cal L}_{\rm gluon}
           +{\cal L}_{\rm quark}+{\cal L}_{\rm ghost}.
\ee
Here ${\cal L}_{\rm massless}$ is a gauge-invariant Lagrangian
for massless gluons and quarks, with null interactions;
${\cal L}_{\rm gluon}$ is the mass and gauge-fixing term
for gluons and auxiliary scalars;
${\cal L}_{\rm quark}$ is the mass term for quarks; and
${\cal L}_{\rm ghost}$ is the Faddeev-Popov ghost term.
The first is given by 
\be
{\cal L}_{\rm massless}=-\frac14\sum_k r_k F_{ak}^{\mu\nu} F_{ak\mu\nu}
  +\sum_i s_i \bar\psi_i i\gamma^\mu\partial_\mu \psi_i  
   +g\sum_{ijk}\beta_i \beta_j \xi_k \bar\psi_i\gamma^\mu T_a A_{ak\mu}\psi_j,
\ee
where the field tensor is
\be
F_{ak}^{\mu\nu}=\partial^\mu A_{ak}^\nu-\partial^\nu A_{ak}^\mu
      -r_k \xi_k gf_{abc}\sum_{lm}\xi_l\xi_m A_{bl}^\mu A_{cm}^\nu.
\ee
The indices are $k$ for (PV) gluons and $i$, $j$ for (PV) quarks,
with values of zero for the physical fields.
The gauge transformations of the fields are
\be
A_{ak}^\mu \longrightarrow  A_{ak}^\mu+\partial^\mu\Lambda_{ak}
+r_k\xi_k g f_{abc}\Lambda_b A_c^\mu,
\ee
\be
\psi_i \longrightarrow \psi_i+ig s_i\beta_i T_a\Lambda_a \psi,
\ee
with $\Lambda_a\equiv\sum_k\xi_k\Lambda_{ak}$ and $[T_a,T_b]=if_{abc}T_c$. 

The regularization is provided by the null field combinations:
\be
A_a^\mu\equiv \sum_k \xi_k A_{ak}^\mu, \;\;
\psi\equiv \sum_i \beta_i \psi_i,
\ee
with $\sum_k r_k \xi_k^2=0$ and $\sum_i s_i \beta_i^2=0$.
Here $r_k$ and $s_i$ equal $\pm1$ ({\em e.g.} $(-1)^k$ and $(-1)^i$),
$A_a^\mu$ is Abelian, and $\psi$ is gauge invariant.
With these definitions, this term of the Lagrangian can be written as
\be \label{eq:Lbasenull}
{\cal L}_{\rm massless}=-\frac14\sum_k r_k(\partial^\mu A_{ak}^\nu-\partial^\nu A_{ak}^\mu)^2
                  +gf_{abc} \partial^\mu A_a^\nu A_{b\mu} A_{c\nu} 
                   +\sum_is_i\bar\psi_i i\gamma^\mu\partial_\mu\psi_i
                  +g\bar\psi\gamma^\mu T_a A_{a\mu}\psi.
\ee
The free terms are those for fields with metrics $r_k$ and $s_i$,
and the interaction terms involve only null fields.
The four-gluon interaction is implicit in the infinite-PV-mass limit
through a contraction of two three-gluon interactions.

The null field combinations provide the necessary two PV subtractions
to regulate any loop.  For example, consider a loop with vertices from
interactions of the form $g\bar\psi\gamma^\mu T_a A_{a\mu}\psi$.
Assume that there is one gluon contraction and one quark
contraction to form the loop.  The contraction of the $k$th gluon 
field yields the metric signature $r_k$ and contraction of the $i$th 
quark field, $s_i$.  The coupling coefficients are selected as 
$\xi_k$, $\xi'_k$, $\beta_i$, and $\beta'_i$.  The loop contribution 
then contains the factors $r_k \xi_k \xi'_k$ and $s_i \beta_i \beta'_i$.
These factors provide the two subtractions, because on their own they
sum to zero

The gluon mass and gauge-fixing term is
\be
{\cal L}_{\rm gluon}=\frac12\sum_k r_k \left(\mu_k A_{ak}^\mu-\partial^\mu\phi_{ak}\right)^2 
   -\frac{\zeta}{2}\sum_k r_k \left(\partial_\mu A_{ak}^\mu +\frac{\mu_k}{\zeta}\phi_{ak}\right)^2.
\ee
The first term is gauge-invariant, and the second, gauge-fixing.
The scalar fields $\phi_{ak}$ obey the gauge transformation
\be
\phi_{ak} \longrightarrow \phi_{ak}+\mu_k \Lambda_{ak}
    +\mu_k r_k \xi_k g f_{abc} \int^x \!\!\!\! dx'_\mu \Lambda_b(x') A_{c}^\mu(x').
\ee
The line integral allows the derivative to transform as
\be
\partial^\mu\phi_{ak} \longrightarrow \partial^\mu\phi_{ak}+\mu_k \partial^\mu\Lambda_{ak}
    +\mu_k r_k \xi_k g f_{abc} \Lambda_b A_{c}^\mu.
\ee
This part of the Lagrangian can be reduced to the form
\be \label{eq:Lgreduced}
{\cal L}_{\rm gluon}=\frac12\sum_k r_k \mu_k^2 \left(A_{ak}^\mu\right)^2
           -\frac{\zeta}{2}\sum_k r_k \left(\partial_\mu A_{ak}^\mu\right)^2 
   +\frac12\sum_k r_k\left[\left(\partial_\mu\phi_{ak}\right)^2
                                  -\frac{\mu_k^2}{\zeta}\phi_{ak}^2\right],
\ee
where the gluon mass $\mu_k$ is explicit and the scalar is given a mass 
$\mu_k/\sqrt\zeta$ and inherits the metric $r_k$.  This is a
non-Abelian extension of a Stueckelberg 
mechanism~\cite{Stueckelberg,Stueckelberg2,Marnelius:1997rx}.

The quark mass term is
\be
{\cal L}_{\rm quark}=\sum_i s_i m_i (\bar\psi_i+ig\frac{s_i\beta_i}{\mu_{\rm PV}}\widetilde\phi_a \bar\psi T_a)
                 (\psi_i-ig\frac{s_i\beta_i}{\mu_{\rm PV}}\widetilde\phi_a T_a\psi),
\ee
with the combination
\be
\widetilde\phi_a\equiv\sum_k \xi_k\frac{\mu_{\rm PV}}{\mu_k}\phi_{ak}
\ee
made null by the additional constraint $\sum_k r_k \frac{\xi_k^2}{\mu_k^2}=0$
and with $\mu_{\rm PV}\equiv \max_k \mu_k$.
The gauge transformation of the combination is Abelian:
\be
\widetilde\phi_a \longrightarrow \widetilde\phi_a+\mu_{\rm PV}\Lambda_a.
\ee

To make all couplings null, we define
\be
\widetilde\psi=\sum_i \beta_i \frac{m_i}{m_{\rm PV}}\psi_i,
\ee
with $m_{\rm PV}\equiv\max_i m_i$,
and impose the constraints $\sum_i s_i m_i^2 \beta_i^2=0$ and
$\sum_i s_i m_i \beta_i^2=0$.  The second constraint makes
$\widetilde\psi$ and $\psi$ mutually null, to allow couplings
between these combinations, and cancels the quartic coupling term.
The quark mass term then simplifies to
\be \label{eq:Lq}
{\cal L}_{\rm quark}=-\sum_i s_i m_i \bar\psi_i\psi_i
            -ig \frac{m_{\rm PV}}{\mu_{\rm PV}}\left[\bar\psi T_a \widetilde\phi_a\widetilde\psi
                                             -\bar{\widetilde\psi} T_a \widetilde\phi_a\psi \right]
\ee

A standard construction~\cite{BailinLove} yields the ghost term~\cite{FaddeevPopov}
\be \label{eq:LFP}
{\cal L}_{\rm ghost}=\sum_k r_k \partial_\mu \bar c_{ak}\partial^\mu c_{ak}
                  -\sum_k r_k \frac{\mu_k^2}{\zeta} \bar c_{ak} c_{ak}
            +g f_{abc}\left[\partial_\mu \bar c_a c_b A_c^\mu
                  -\frac{\mu_{\rm PV}^2}{\zeta}\bar{\widetilde c}_a
                       \int^x\!\!\!\! dx'_\mu c_b(x') A_c^\mu(x')\right],
\ee
for ghosts $c_{ak}$ and anti-ghosts $\bar c_{ak}$,
with null combinations defined as
\be
c_a\equiv\sum_k\xi_k c_{ak}, \;\;
\bar c_a\equiv\sum_k\xi_k \bar c_{ak}, \;\;
\bar{\widetilde c}_a\equiv\sum_k \xi_k\frac{\mu_k^2}{\mu_{\rm PV}^2}\bar c_{ak}.
\ee
For these to be (mutually) null, we require
$\sum_k r_k \mu_k^2 \xi_k^2=0$ and 
$\sum_k r_k \mu_k^4 \xi_k^2=0$.

As a summary of the various constraints, the following need to be satisfied
for for the adjoint fields: 
\be
\sum_k r_k \xi_k^2=0, \;\;
\sum_k r_k \frac{\xi_k^2}{\mu_k^2}=0, \;\;
\sum_k r_k \mu_k^2 \xi_k^2=0, \;\;
\sum_k r_k \mu_k^4 \xi_k^2=0,
\ee
and for the quark fields: 
\be
\sum_i s_i \beta_i^2=0,  \;\;
\sum_i s_i m_i^2 \beta_i^2=0, \;\;
\sum_i s_i m_i \beta_i^2=0.
\ee
For the PV masses to be chosen independently, these constraints
require four PV gluons, four PV ghosts and antighosts, five PV 
scalars, and three PV quarks.  For pure Yang--Mills theory, the 
number of PV scalars is four, with the $k=0$ field dropped and 
$\mu_0=0$; the fields $\phi_{a0}$ are used only in splitting the 
masses of the PV quarks.  In either case, the number of PV fields
translates to a large, but necessary computational load.

The BRST transformations are
\bea
\delta A_{ak}^\mu&=&\epsilon\partial^\mu c_{ak}
                      +\epsilon r_k\xi_k g f_{abc}c_b A_c^\mu, \\
\delta \psi_i &=& i\epsilon g s_i \beta_i T_a c_a \psi, \;\;
\delta \bar\psi_i = -i\epsilon g s_i \beta_i \bar\psi T_a c_a, \\
\delta \phi_{ak}&=&\epsilon\mu_k c_{ak} +\epsilon r_k\xi_k \mu_k g f_{abc}
                             \int^x\!\!\!\! dx'_\mu c_b(x') A_c^\mu(x'), \\
\delta \partial^\mu\phi_{ak}&=&\epsilon\mu_k \partial^\mu c_{ak}
                      +\epsilon r_k\xi_k \mu_k g f_{abc} c_b A_c^\mu, \\
\delta \bar c_{ak} & =& -\zeta\epsilon\left(\partial_\mu A_{ak}^\mu+\frac{\mu_k}{\zeta}\phi_{ak}\right), \;\;
\delta c_{ak} = \frac12 \epsilon r_k \xi_k g f_{abc}c_b c_c,
\eea
with $\epsilon$ a real Grassmann constant, for which $\epsilon^2=0$.
For the various null combinations, we then find
\be
\delta A_a^\mu = \epsilon\partial^\mu c_a, \;\;
\delta \widetilde\phi_a= \epsilon\mu_{\rm PV}c_a, \;\;
\delta c_a = 0, \;\;
\delta \psi =0, \;\;
\delta \widetilde\psi = 0.
\ee
The full Lagrangian is invariant with respect to these transformations.

%%%%%%%%%%%%%%%%%%%%%%%%%%%%%%%%%%%%%%%%%%%%%
\section{Massive vector quantization}
%%%%%%%%%%%%%%%%%%%%%%%%%%%%%%%%%%%%%%%%%

The construction of the Lagrangian makes no reference to a choice
of quantization coordinates; however, the construction was done
in such a way as to keep open the possibility of light-front
quantization in an arbitrary covariant gauge.  To see how this
quantization can take place, consider the case of QED~\cite{ArbGauge}.
The Lagrangian is ${\cal L}=-\frac14 F^2+\frac12\mu A^2-\frac12\zeta(\partial\cdot A)^2$,
which yields the field equation: $(\Box +\mu^2)A_\mu-(1-\zeta)\partial_\mu(\partial\cdot A)=0$.
The light-front Hamiltonian density is
\be
{\cal H}={\cal H}|_{\zeta=1}
  +\frac12(1-\zeta)(\partial\cdot A)(\partial\cdot A
                                -2\partial_-A^+-2\partial_\perp\cdot\vec A_\perp),
\ee
with
\be
{\cal H}|_{\zeta=1}=\frac12\sum_{\mu=0}^3 \epsilon^\mu
       \left[(\partial_\perp A^\mu)^2+\mu^2 (A^\mu)^2\right]
\ee
and $\epsilon^\mu=(-1,1,1,1)$.  From this we obtain the
light-front Hamiltonian
\be
{\cal P}^-=\int d\ub{x}{\cal H}|_{x^+=0}
  =\int d\ub{k} \sum_\lambda \epsilon^\lambda \frac{k_\perp^2+\mu^{(\lambda)2}}{k^+}
                a^{(\lambda)\dagger}(\ub{k})a^{(\lambda)}(\ub{k})
\ee
with $\ub{k}=(k^+,\vec k_\perp)$, $\mu^{(\lambda)}=\mu$ for $\lambda=1,2,3$, 
but $\mu^{(0)}=\tilde\mu\equiv\mu/\sqrt{\zeta}$.
The nonzero commutator is 
\be
[a^{(\lambda)}(\ub{k}),a^{(\lambda')\dagger}(\ub{k}')]
     =\epsilon^\lambda \delta_{\lambda\lambda'}\delta(\ub{k}-\ub{k}').
\ee

The normal-mode expansion for the field is
\be
A_\mu(x)=\int\frac{d\ub{k}}{\sqrt{16\pi^2 k^+}}\left\{\sum_{\lambda=1}^3
   e_\mu^{(\lambda)}(\ub{k})\left[ a^{(\lambda)}(\ub{k})e^{-ik\cdot x}
            + a^{(\lambda)\dagger}(\ub{k})e^{ik\cdot x}\right] 
    +e_\mu^{(0)}(\ub{k})\left[ a^{(0)}(\ub{k})e^{-i\tilde k\cdot x}
            + a^{(0)\dagger}(\ub{k})e^{i\tilde k\cdot x}\right]\right\},
\ee
with polarization vectors
\be
e^{(1,2)}(\ub{k})=(0,2 \hat e_{1,2}\cdot \vec{k}_\perp/k^+,\hat e_{1,2}), \;\;
e^{(3)}(\ub{k})=\frac1\mu((k_\perp^2-\mu^2)/k^+,k^+,\vec k_\perp), \;\;
e^{(0)}(\ub{k})=\tilde k/\mu.
\ee
Here $\ub{\tilde k}=\ub{k}$, $\tilde k^-=(k_\perp^2+\tilde\mu^2)/k^+$,
and $\hat e_{1,2}$ are transverse unit vectors.
These vectors satisfy: $k\cdot e^{(\lambda)}=0$ and 
$e^{(\lambda)}\cdot e^{(\lambda')}=-\delta_{\lambda\lambda'}$
for $\lambda,\lambda'=1,2,3$.
The first term in $A_\mu$ satisfies $(\Box +\mu^2)A_\mu=0$ and
$\partial\cdot A=0$.  The $\lambda=0$ term violates each, but the
field equation is satisfied~\cite{Stueckelberg,Coester}.

The nondynamical components of the fermion fields satisfy the
constraints
\be \label{eq:FermionConstraint}
is_i\partial_-\psi_{i-}+e A_-\beta_i\sum_j\psi_{j-}
  =(i\gamma^0\gamma^\perp)
     \left[s_i\partial_\perp \psi_{i+}-ie A_\perp\beta_i\sum_j\psi_{j+}\right] 
   -s_i m_i \gamma^0\psi_{i+}.
\ee
When multiplied by $s_i\beta_i$ and summed over $i$,
this becomes
\be
i\partial_-\psi_-
  =(i\gamma^0\gamma^\perp)
     \partial_\perp \psi_+
      - \gamma^0\sum_i\beta_i m_i\psi_{i+},
\ee
which is the same constraint as for a free fermion field, in any gauge.
Quantization can then proceed without need of light-cone gauge.

%%%%%%%%%%%%%%%%%%%%%%%%%%%%%%%%%%%%%%%%%%%%%%%
\section{Light-front coupled-cluster method}
%%%%%%%%%%%%%%%%%%%%%%%%%%%%%%%%%%%%%%%%%%%%%

Given a field-theoretic light-front Hamiltonian $\Pminus$, we wish
to solve the fundamental eigenvalue problem
$\Pminus|\psi\rangle=\frac{M^2+P_\perp^2}{P^+}|\psi\rangle$.
The LFCC method~\cite{LFCClett} is designed to do so in terms of a Fock-state
expansion but without the usual truncation of Fock space.
Instead, write the eigenstate as $|\psi\rangle=\sqrt{Z}e^T|\phi\rangle$.
Here $Z$ controls the normalization, which is fixed as
 $\langle\psi'|\psi\rangle=\delta(\ub{P}'-\ub{P})$.
The ket $|\phi\rangle$ is the valence state, with norm
   $\langle\phi'|\phi\rangle=\delta(\ub{P}'-\ub{P})$.
The operator $T$ contains terms that only increase particle number;
it does, however, conserve the quantum numbers of the full state,
such as $J_z$, light-front momentum $\ub{P}$, and charge.  Because
$p^+$ is always positive, $T$ must include annihilation operators,
and powers of $T$ include contractions. 
With construction of the effective Hamiltonian $\ob{{\cal P}-}=e^{-T} \Pminus e^T$
and the definition of a projection $P_v$ onto the valence Fock sector,
we have the coupled system:
\be
P_v\ob{\Pminus}|\phi\rangle=\frac{M^2+P_\perp^2}{P^+}|\phi\rangle, \;\;
(1-P_v)\ob{\Pminus}|\phi\rangle=0.
\ee

This is still an infinite system of equations.  The LFCC approximation
is to truncate $T$ at a fixed increase in particle count, but
not truncate the exponential of $T$.  The projection $(1-P_v)$ is
truncated to a consistent set of Fock sectors above the valence sector,
enough to have a finite, but sufficient system of (nonlinear) equations 
for the functions in $T$ and in the valence state.
The leading approximation of $T$ for QCD
would include quark $\rightarrow$ quark + gluon,
gluon $\rightarrow$ quark + antiquark, and
gluon $\rightarrow$ gluon + gluon transitions.

The LFCC approach requires no Fock-space truncation and
no sector dependence or spectator dependence of self-energies
or bare parameters.  It is systematically improvable by
adding terms to $T$ and sectors to $(1-P_v)$ with more 
constituents.  Various applications have been carried out,
including a heavy-fermion model~\cite{LFCClett},
$\phi^4$ theory~\cite{LFCCphi4}, and QED~\cite{LFCCqed}.

%%%%%%%%%%%%%%%%%%%%%%%%%%%%%%%%%%%%%%%%%%%%%%%%%%%%%%%%%%
\section{Summary}  \label{sec:summary}
%%%%%%%%%%%%%%%%%%%%%%%%%%%%%%%%%%%%%%%%%%%%%%%%%

With these tools in place, the time has come for a
concentrated assault on QCD.  The procedure would be
to invoke PV regularization and light-front quantization
in an arbitrary covariant gauge with variable gauge parameter.
Any (numerical) approximation~\cite{Vary,Glazek,Karmanov}
would then be applied to a
finite theory where the continuum limit can be taken
independent of any chosen renormalization scheme.
Fock-space truncation and the resulting uncanceled
divergences can be avoided by using the LFCC method.
The best places to start would be simpler systems in 
quenched approximation, such as heavy-quark mesons
and glueballs.  The computational load and effort 
are likely comparable to lattice gauge theory;
the large number of PV modes and the discretizations
being the analogs of lattice size and spacing.

%%%%%%%%%%%%%%%%%%%%%%%%%%%%%%%%%%%%%%%%%%%%%%%%%%
\acknowledgments
%%%%%%%%%%%%%%%%%%%%%%%%%%%%%%%%%%%%%%%%%%%%%%%%%%
This work was done in collaboration with S.S. Chabysheva.
Travel to the conference was supported in part by the
University of Minnesota Global Programs \& Strategy Alliance.

\end{document}